\documentclass[floatfix,superscriptaddress,a4paper,
               nofootinbib,notitlepage,12pt]{revtex4-2}

\usepackage[colorlinks=true,linktocpage=true,linkcolor=blue,citecolor=blue,allcolors=blue]{hyperref}
\pdfoutput=1

\usepackage{graphicx}
\usepackage{amsmath}
\usepackage{amssymb}
\usepackage{bm}

\newcommand{\hsp}{\hspace*{1pt}}
\newcommand{\hspm}{\hspace*{.5pt}}
\newcommand{\ds}{\displaystyle}
\newcommand{\be}{\begin{equation}}
\newcommand{\ee}{\end{equation}}
\newcommand{\bel}[1]{\be\label{#1}}
\newcommand{\re}[1]{Eq.~(\ref{#1})}

\usepackage{color}

\begin{document}

\title
{
Phase diagram of bosonic matter\\
with additional derivative interaction
}

\author{L.~M. Satarov}
\affiliation{
Frankfurt Institute for Advanced Studies, D-60438 Frankfurt am Main, Germany}

\author{I.~N. Mishustin}
\affiliation{
Frankfurt Institute for Advanced Studies, D-60438 Frankfurt am Main, Germany}

\author{H. Stoecker}
\affiliation{
Frankfurt Institute for Advanced Studies, D-60438 Frankfurt am Main, Germany}

\begin{abstract}
{Equation of state of uncharged bosonic matter (BM)
is studied within a~field-theoretical approach in the mean-field approximation.
Interaction of bosons is described by a scalar field~$\sigma$ with
a~Skyrme-like potential which contains both attractive and repulsive terms.
Additionally we introduce the derivative
interaction~(DI) by including factor $(1+\lambda\sigma)^{-1}$ in the kinetic part
of Lagrangian where $\lambda > 0$ is the model parameter.
Numerical calculations are made for strongly interacting matter
composed of $\alpha$ particles.
It is shown that ground-state binding energy and equilibrium density
of such matter drop with increasing $\lambda$. The liquid-gas
phase transition (LGPT) and the Bose-Einstein condensation (BEC) are studied
by using different thermodynamic variables. We calculate also
spinodal lines which give boundaries of metastable states.
It is demonstrated that critical temperature decreases 
with~$\lambda$. Both LGPT and bound condensate states disappear above certain
maximum value~of~$\lambda$.}
\end{abstract}


 \maketitle

\section{Introduction}

The derivative (momentum--dependent) interactions are widely used in modern nuclear and particle physics.
They are introduced e.g. for implementing the effects of binary scatterings with nonzero angular
momenta ($p,d...$ waves), for accounting finite sizes of hadrons, for analyzing surface properties of nuclei  etc.
One example of such interaction is the pion-nucleus optical potential of the Kislinger type~\cite{Kis74}.
Inclusion of additional gradient terms of the potential lead to better agreement with observed data on $\pi A$ elastic scattering.
The phenomenological Skyrme potentials of~NN interactions ~\cite{Sky56,Vau72} also contain momentum-dependent terms.
In~addition, one should mention the relativistic mean--field models~\cite{Zim90,Che14} with derivative couplings of fermion and meson fields.
By using such interactions one can achieve reasonable values of the nuclear matter compressibility, overestimated in standard
Walecka calculations.

A large variety of models with DI have been formulated in cosmology.
For example, the 'kinetic K-essence' mo\-dels~\mbox{\cite{Mel03,Sch04}} with noncanonical kinetic energies
of scalar fields were introduced to explain the Universe acceleration, observed in red--shift measurements
of distant supernovae. The authors of Ref.~\cite{Llo22} introduce the coupling between
the cosmic curvature and the density of the scalar field $\sigma$. This leads to the kinetic part of matter Lagrangian
with an additional factor $F=(1+A\sigma)^{-1}$. Below we use this result to investigate~BEC and phase transition
of bosonic matter with DI.

In Refs.~\cite{Sat17,Sat21} we developed the mean--filed models to describe properties of homogeneous matter composed of $\alpha$ particles.
We use the standard standard kinetic energy and describe interactions of particles by using the Skyrme potential which is attractive
at small densities. Similarly to atomic $^4$He, such matter undergoes both the LGPT and the BEC transition. In Refs.~\cite{Sat19,Sat20} we
have studied the $\alpha$-nucleon matter.  It is shown that at large enough attraction between $\alpha$ particles and nucleons
the nuclear ground state contains the admixture of $\alpha$'s. A similar effect has been predicted within the nuclear lattice
model in Ref.~\cite{Elh16}.

In our recent publication~\cite{Sat22} we studied characteristics of '$\alpha$-conjugate' nuclei (with even proton
and neutron numbers $Z=N=A/2$)\,. It is believed that such nuclei may contain Bose-condensed clusters of $\alpha$
particles (see Ref.~\cite{Oer10} and references therein). We have considered ground states of these nuclei
as Q--balls\hspm\footnote
{
The 'Q--balls' have been introduced in Refs.~\cite{Ros68,Fri76,Col85}  as soliton-like bound states of scalar bosons. Currently they are considered as promising
candidates for stable dark matter.
}
(or Q--shells) made of charged $\alpha$ particles. It was shown that typically such Q--balls have bindings which noticeable overestimate the observed
binding energies of $\alpha$-conjugate' nuclei. In addition, the standard calculation predicts too small surface tension which leads to
unrealistically narrow widths\hspm\footnote
{
less than the size of the $\alpha$ particle
}
of Q-ball surfaces. This unsatisfactory result has been corrected in Ref.~\cite{Sat22} by introducing
an artificially enhanced gradient term of the energy density. We believe that future covariant calculations with derivative interaction may resolve this problem.

In the present article we study the sensitivity of the phase diagram of bosonic matter to~DI.
As far as we know, this has not been done up to now by other authors. We consider the particular case of matter composed of $\alpha$ particles
and use the parameters of their potential from our previous publications~\cite{Sat21,Sat22}.

The paper is organized as follows. In Sec.~\ref{eom} we derive the generalized Klein-Gordon equation for the bosonic field.
It is obtained using the kinetic term of the Lagrangian with the derivative factor. In the next section we consider the thermodynamic
functions for cold~BM. In Sec.~\ref{numr} we calculate numerically properties of BM at zero temperature for different
sets of model parameters. In Sec.~\ref{nzt} we extend our formalism to nonzero temperatures. The resulting
formulae are applied in Sec.~\ref{phd} for investigating the phase diagram of warm BM.

\section{Cold bosonic matter in the mean-field approximation}
\label{sec-model}

\subsection{Equations of motion for scalar field~\label{eom}}

In this paper we consider a system of uncharged bosonic particles with strong self-interaction.
We apply a field-theoretical approach denoting by $\phi\hspm (x)$ the scalar field at {the} space-time point $x^\nu=(t,\bm{r})^\nu$.
 The  Lagrangian density {of the} system is assumed to be of the generic form ($\hbar=c=1$):
 \bel{lagd0}
{\cal L}=\frac{1}{2}\hspm F(\sigma)\chi - U(\sigma)\,,
\ee
where \mbox{$\sigma=\phi\hsp\phi^*$}, \mbox{$\chi=\partial_\nu\hsp\phi\,\partial^\nu\hspace{-.5ex}\phi^*$} and $U(\sigma)$ is
the mean-field potential, {which contains the mass term and }
self-interaction terms. One can regard the quantity $[F(\sigma)-1]\chi/2$
as an additional~DI proportional to particle four-momentum squared.
In the majority of theoretical models it is implicitly
assumed that $F=1$ i.e. one neglects any modification of the kinetic part of Lagrangian
as compared to its canonical form ${\cal L}_{\rm kin}=\chi/2$\,.

By using~\re{lagd0} one can write the expressions for the four current $J^{\nu}$ and the energy-momentum
tensor $T^{\mu\nu}$:
\bel{cen}
J^{\nu}=\frac{iF}{2}\left(\phi\,\partial^{\,\nu}\hspace{-.4ex}\phi^*-\phi^*\partial^{\,\nu}\hspace{-.4ex}\phi\right),~~
T^{\mu\nu}=\frac{F}{2}\left(\partial^{\,\mu}\hspace{-.4ex}\phi\,\partial^{\,\nu}\hspace{-.4ex}\phi^*+
\partial^{\,\mu}\hspace{-.5ex}\phi^*\partial^{\,\nu}\hspace{-.5ex}\phi\right)-g^{\mu\nu}{\cal L}\,.
\ee
The Euler-Lagrange equation leads to the equation of motion for scalar fields.
One gets the generalized Klein-Gordon equation (KGE)
\bel{ele}
\partial_\nu (F\partial^{\,\nu}\hspace{-.4ex}\phi)+(2\,U^{\hspm\prime} -F^{\hsp\prime}\chi)\hsp\phi=0\,,
\ee
where primes denote derivatives with respect to $\sigma$\,.

The stationary solution
of the KGE is obtained by the substitution~\cite{Col85}
\bel{ssol}
\phi=e^{i\mu t}\varphi\hsp (\bm{r})\,,
\ee
where $\varphi=\sqrt{\sigma}$ is a real function of spatial coordinates $\bm{r}$.
The constant $\mu$ is in fact the chemical potential characterizing
the bound state of the Bose-Einstein condensate at zero temperature (see below).
Substituting (\ref{ssol}) into \re{ele} one gets the stationary KGE for $\varphi$\hsp\footnote
{
In case $F=1$ one obtains the equation for the
Q-ball profile from~Ref.~\cite{Col85}.
}
\bel{sele}
{\bm\nabla}(F\hsp {\bm\nabla}\varphi)+\left[\mu^2\,(F\sigma)^\prime-2\hspm U^{\,\prime}-F^{\hsp\prime}\hsp ({\bm\nabla}\varphi)^{\hspm 2}\right]\varphi=0\hsp .
\ee

Using (\ref{ssol}) in Eqs.~(\ref{cen}) gives the expressions
for the number density $n$ and the energy density~$\varepsilon$ of
bosonic particles
\begin{eqnarray}
&&n=J^{\hsp 0}=\mu F\sigma\,,~ \label{dens}\\
&&\varepsilon=T^{\hsp 00}=\frac{\mu\hsp n}{2}+U+\frac{F}{2}\left({\bm\nabla}\varphi\right)^{\hspm 2}.~ \label{edens}
\end{eqnarray}

It is interesting that \re{sele} can be also obtained by minimizing the functional
of thermodynamic potential $\Omega=\varepsilon-\mu n$ with respect to arbitrary
variation $\delta\phi$ at constant $\mu$\,. One gets the relation
\bel{minf}
\int dV \delta\hsp\Omega=\int dV \delta\hspace{-.3ex}\left[\frac{F}{2}\left({\bm\nabla}\varphi\right)^{\hspm 2}
-p\hsp (\mu,\sigma)\right]=0\,.
\ee
where
\bel{prem}
p\hsp (\mu,\sigma)=\frac{\mu^2\hsp F\sigma}{2}-U(\sigma)\,.
\ee

Let us consider now the homogeneous BM at zero temperature.
In this case, substituting ${\bm\nabla}\varphi=0$,
we get the 'gap' equation determining the equilibrium condensate density
$\sigma=\sigma (\mu)$ and pressure $p$ at given $\mu$
\bel{gap}
\left(\frac{\partial\hsp p}{\partial\hsp\sigma}\right)_{\hspace*{-1pt}\mu}=
\frac{\mu^2 (F\hsp\sigma)^{\hsp\prime}}{2}-U^{\hsp\prime} (\sigma)=0\,.
\ee
Using Eqs.~(\ref{dens}),~(\ref{prem})--(\ref{gap}) one
can prove the condition of thermodynamic consis\-ten\-cy~\mbox{$dp=n\hsp d\mu$}.
It is convenient to rewrite \re{gap} in the form
\bel{gap1}
\mu=M(\sigma),~~M\hsp (\sigma)=\sqrt{\dfrac{2\hsp U^\prime(\sigma)}{(F\hspm\sigma)^\prime}}\,.
\ee
As we will see below $M\hsp (\sigma)$ plays a role of the effective mass of
bosonic quasiparticles. The first relation in (\ref{gap1}) may be considered as
condition of BEC.

\subsection{Thermodynamic functions and ground states at $T=0$\label{st0}}

At thermal equilibrium all particles in cold BM are in the Bose-condensed state with zero momentum (within the mean-field approximation).
We introduce the following parametrization of functions $F$\hspm\footnote
{
It is interesting that the same parametrization of the derivative factor
has been used in the cosmological model~of Ref.~\cite{Llo22}.
}
and $U$:
\bel{fup}
F (\sigma) = (1+\lambda\hsp\sigma)^{-1},~~U (\sigma) = \frac{\sigma}{2}\left(m^2-\frac{a\hsp \sigma}{2}+\frac{b\hsp\sigma^{\alpha}}{\alpha+1}\right),
\ee
where $m$ is the vacuum mass of bosons and $a,b,\lambda$ and $\alpha$ are positive model parameters~(we consider only the values $\alpha>1$).
Note that the second and third terms of Skyrme potential~$U (\sigma$) describe respectively the
attractive and repulsive interactions of particle.
The~limit~$\lambda=0$ corresponds to the case
without DI. It was considered in our previous publications on pionic~\cite{Mis19} and
$\alpha$-particle~\cite{Sat17,Sat21,Sat22} systems\hsp\footnote
{
Note that only the case $\alpha=2$ has been considered in Refs.~\cite{Mis19,Sat21,Sat22}.
}.
Below we show that the presence
of the suppression factor $F$ leads effectively to an additional repulsion.

Using formulae of preceding section, one gets the equations
\bel{mum}
\mu=M(\sigma)=(1+\lambda\hsp\sigma)\hsp\sqrt{m^2-a\hspm\sigma+b\hsp\sigma^\alpha}\,,
\ee
determining the scalar density $\sigma$ and the effective mass~$M$
as functions of the chemical potential~$\mu$\hsp\footnote
{
In general, \re{mum}
has several solutions for $\sigma=\sigma (\mu)$\,.
In real calculations we use $\sigma$ as input variable
and determine $\mu$ as function of $\sigma$\,.
}. According to~\re{mum} characteristics of the Bose condensate
are sensitive to~$\lambda$\,. One can see that at fixed $\sigma$ increasing $\lambda$ leads
to increase of the effective mass and therefore, to reducing
binding energy $m-M$\hsp . As a consequence,
the LGPT lines, appearing in the present model, move with growing
$\lambda$ to the region of larger chemical potentials (see below~Figs.~\ref{fig6}--\ref{fig8}).

Finally, we get the following equations for
density $n$, pressure $p$ and energy density $\varepsilon$:
\begin{eqnarray}
&&n=\sigma\hspm\sqrt{m^2-a\hspm\sigma+b\hspm\sigma^\alpha},\label{den}\\
&&p=\dfrac{M\hspm n}{2}-U\hsp,~\varepsilon=\dfrac{M\hspm n}{2}+U\,.\label{pre0}
\end{eqnarray}

Let us consider properties of the ground state (GS) of cold BM, i.e.
the state of minimum energy per particle or zero pressure. Using the above equations
one can determine the ground-state scalar density $\sigma_0$ and the corresponding chemical
potential $\mu_{\hspm 0}$ by solving the system of equations
\bel{gsp}
\mu_0=(1+\lambda\hsp\sigma_0)\sqrt{2\hspm U^{\hsp\prime}_0}\,,
~~p_0=\frac{\mu_0^2\hsp\sigma_0}{2\hsp (1+\lambda\hsp\sigma_0)}-U_0=0\,,
\ee
where quantities with index zero are taken at $\sigma=\sigma_0$\,.
Eliminating the chemical potential gives the relation
\bel{gsp1}
1+\lambda\hsp\sigma_0=\frac{U_0}{\sigma_0\hspm U^{\hsp\prime}_0}\,,
\ee
which determines the dependence of $\sigma_0$ on the parameter $\lambda$.
Using further the explicit form of~$U\hsp (\sigma_0)$ from \re{fup}
we get the expression
\bel{gsp2}
\lambda\hsp (m^2-a\hspm\sigma_0+b\hspm\sigma_0^\alpha)=\frac{a}{2}-\frac{\alpha\hspm b\hspm\sigma_0^{\alpha-1}}{\alpha+1}\,,
\ee
which was solved numerically at fixed parameters $a,b,\lambda$\,.

In the limit $\lambda\to 0$ characteristics of GS
can be found analytically for any $\alpha$\,. Below we mark
the corresponding values by tilde. Substituting
$\lambda=0$ in Eqs. (\ref{gsp}), (\ref{gsp2})
leads to the relation
\bel{s0t}
\tilde{\sigma}_0=
\left[\frac{(1+\alpha)\hsp a}
{2\hspm\alpha\hsp b}\right]^{\frac{\ds 1}{\ds\alpha-1}}.
\ee
In the present paper we chose $a,b$ by fixing characteristics
of the GS at zero derivative strength $\lambda$\hspm .
Namely, we assume the same binding energy per particle
$W_0=m-\tilde{\mu}_0$ and density $n_0$
for all $\alpha$.
One can write down the explicit relations
\begin{eqnarray}
&&a=\dfrac{2\hspm\alpha}{\alpha-1}\dfrac{(m^2-\tilde{\mu}_0^2)}{\tilde{\sigma}_0}\,, \label{eqa}\\
&&b=\dfrac{\alpha+1}{\alpha-1}\dfrac{(m^2-\tilde{\mu}_0^2)}{\tilde{\sigma}_0^\alpha}\,,\label{eqb}
\end{eqnarray}
where $\tilde{\sigma}_0=n_0/\tilde{\mu}_0$\,.

As an example, we use the GS characteristics
determined in variational calculations of cold~$\alpha$ matter
in Ref.~\cite{Cla66}\hsp\footnote
{
These calculations were based on phenomenological pair potentials
adjusted to $\alpha\alpha$ phase shifts observed at low energies.
}:
\bel{clav}
W_0=19.7~\textrm{MeV}, n_0=0.036~\textrm{fm}^{-3}\,.
\ee
Using this procedure and substituting the particle
mass \mbox{$m=m_\alpha=3.7273~\textrm{GeV}$}
we get numerical values of $a,b$ for two parameter sets with  $\alpha=1.1$ and
$\alpha=2$ (see Table~\ref{tab1}). Note that both these sets
correspond to the same value of scalar density
\bel{s0t1}
\tilde{\sigma}_0\simeq 9.658\cdot 10^{-3}~\textrm{GeV}^{-1}\hsp\textrm{fm}^{-3}\,.
\ee

The case of nonzero $\lambda$ can be discussed qualitatively in
the nonrelativistic approxima\-tion~(NRA). It has a good accuracy, at least for $\alpha$
matter at not too large scalar densities. Within the NRA $|\mu_0-m|\ll m$\,,
and one can omit the terms containing $\sigma_0$ in the left hand side of~\re{gsp2}.
This leads to the approximate formulae
\bel{gsp3}
\sigma_0\simeq\tilde{\sigma}_0\left(1-\frac{\lambda}{\lambda_{\rm m}}\right)^{\frac{\ds 1}{\ds\alpha-1}}\,,
\ee
where $\tilde{\sigma}_0$ is defined in (\ref{s0t}) and $\lambda_{\rm m}=a/(2\hsp m^2)$ is the maximal possible value of $\lambda$
at which bound states of BM exist~($\mu_0<m$)\,. One can see that $\sigma_0$ monotonically drops
with $\lambda$\,.
The last column of Table~\ref{tab1} gives values of $\lambda_{\rm m}$ for
two choices of $\alpha$ considered in this paper. One can see that harder Skyrme potentials
(with larger $\alpha$) lead to reduced values of~$\lambda_m$\hsp\footnote
{
Using the data of Table~\ref{tab1} one can show that the factor $F$
deviates only slightly from unity at $\sigma\lesssim\sigma_0$. Nevertheless,
the derivative corrections have a relatively large effect even in NRA.
This follows from the fact that zero-order terms in $\lambda$
(originating from large mass term of Skyrme potential)
effectively cancel out in pressure and binding energy (see Eqs. (\ref{gsp})--(\ref{gsp2}))\,.
}.

\begin{table}[ht!]
\caption
{\label{tab1} Parameters of Skyrme potential and maximal value of derivative
strength $\lambda_{\rm m}$\,.}
\vspace*{3mm}
\begin{tabular}{|c|c|c|c|}
\hline
~$\alpha$~&~$a\,\left(\textrm{GeV}^{\,3}\hsp\textrm{fm}^3\hsp\right)$~&
~$b\,\left(\textrm{GeV}^{\,\alpha+2}\hsp\textrm{fm}^{\hsp 3\hspm\alpha}\hsp\right)$~&~$\lambda_{\rm m}\,\left(\textrm{GeV}\hspm\textrm{fm}^{3}\hsp\right)$~\\
\hline
~1.1~~& 331.9& 503.5 & 11.9 \\
\hline
2~~& 60.34 & 4661 & 2.16 \\
\hline
\end{tabular}
\end{table}

Within the considered model one can easily calculate the sound velocity $c_s$. At $T=0$ one may
use the expressions~\cite{Lan75}
\bel{svel}
c_s^{\,2}=\dfrac{d\hsp p}{d\hsp\varepsilon}=\frac{n\hsp d\mu}{\mu\hsp dn}\,.
\ee
Substituting $\mu=M\hsp (\sigma), n=M\hsp F\sigma$ and using the relation $(F\hsp\sigma)^\prime=F^2$
we obtain the equation
\bel{svel1}
c_s^{\,2}=\left[1+\frac{M F}{M^\prime\sigma}\right]^{-1}.
\ee
One can see that unstable spinodal states with $c_s^2<0$ occur at $1+M F/M^\prime\sigma <0$. Within the
NRA one obtains
\bel{svel2}
c_s^{\,2}\simeq\frac{M^\prime (\sigma)\hsp\sigma}{m\hspm F(\sigma)}\,.
\ee

\subsection{Numerical results\label{numr}}

In this section we present results of numerical calculations
for cold BM obtained within the scheme described in
preceding section. We apply the model parameters given in Table~\ref{tab1}\,.
Table~\ref{tab2} shows the characteristics of GS calculated
for different $\alpha$ and $\lambda$\,. One can see that
the ground-state binding energy, density and sound velocity of BM are
rather sensitive to the derivative strength $\lambda$. According to our calculations,
all these quantities drop with increasing  $\lambda$.
\begin{table}[ht!]
\caption
{\label{tab2} Ground-state characteristics of BM at $T=0$\,.}
\vspace*{3mm}
\begin{tabular}{|c|c|c|c|c|c|c|}
\hline
&\multicolumn{3}{c|}{$\alpha=1.1$}&\multicolumn{3}{c|}{$\alpha=2$}\\
\cline{2-7}
$\lambda$ &~$\mu_0-m$~&~$n_0\times 10^2$ &~$c_s\times 10^2$ &~$\mu_0-m$ &~$n_0\times 10^2$ &~$c_s\times 10^2$  \\[-1.5mm]
$\left(\textrm{GeV}\hsp\textrm{fm}^{3}\right)$~&~$\left(\textrm{MeV}\right)$~&~$\left(\textrm{fm}^{-3}\right)$ &~&~$\left(\textrm{MeV}\right)$
&~$\left(\textrm{fm}^{-3}\right)$ &~\\
\hline
$0$ & $-19.70$ & $3.60$ & $7.63$ & $-19.70$ & $3.60$ & $10.3$\\
\hline
$0.5$ & $-12.35$ & $2.36$ & $6.07$ & $-11.73$ & $2.78$ & $7.96$ \\
\hline
$1$ & $-7.58$ & $1.96$ & $4.76$ & $-5.81$ & $1.52$ & $5.63$\\
\hline
\end{tabular}
\end{table}

\begin{figure}[htb!]
\centering
\includegraphics[trim=2cm 7.5cm 3cm 7.5cm,width=0.6\textwidth]{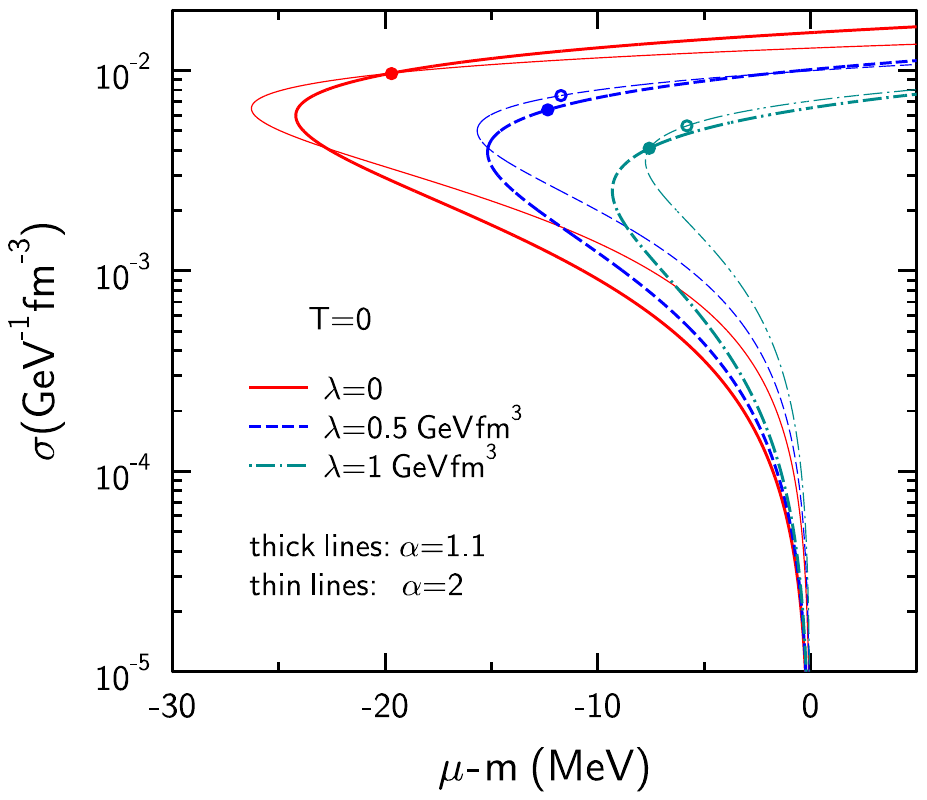}
\caption{
The scalar density of cold BM as the function of chemical potential $\mu$
for different values of derivative strength $\lambda$\,.
Bold and thin lines are calculated from~\re{mum} with $\alpha=1.1$
and $\alpha=2$\,, respectively. Dots show positions
of GS on the ($\mu,\sigma$) plane.
}\label{fig1}
\end{figure}
\begin{figure}[hb!]
\centering
\includegraphics[trim=2cm 7.5cm 3cm 7.5cm,width=0.6\textwidth]{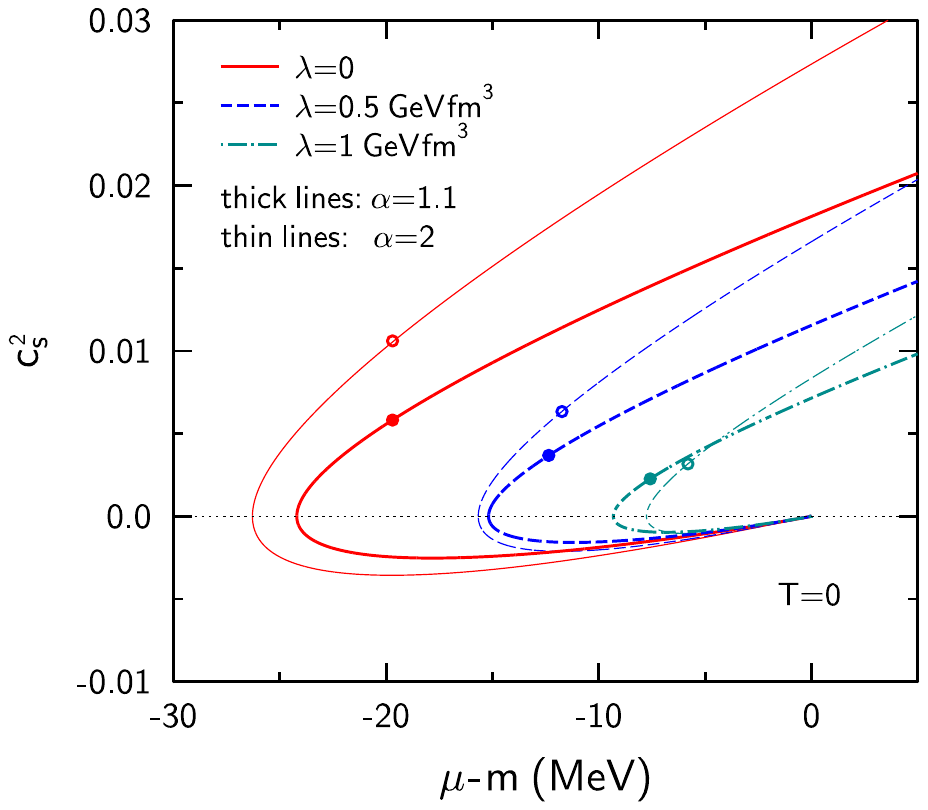}
\caption{
Same as Fig.~\ref{fig1}, but for the sound velocity squared as the function of chemical
potential.
}\label{fig2}
\end{figure}
\begin{figure}[htb!]
\centering
\includegraphics[trim=2cm 7.5cm 3cm 8cm,width=0.6\textwidth]{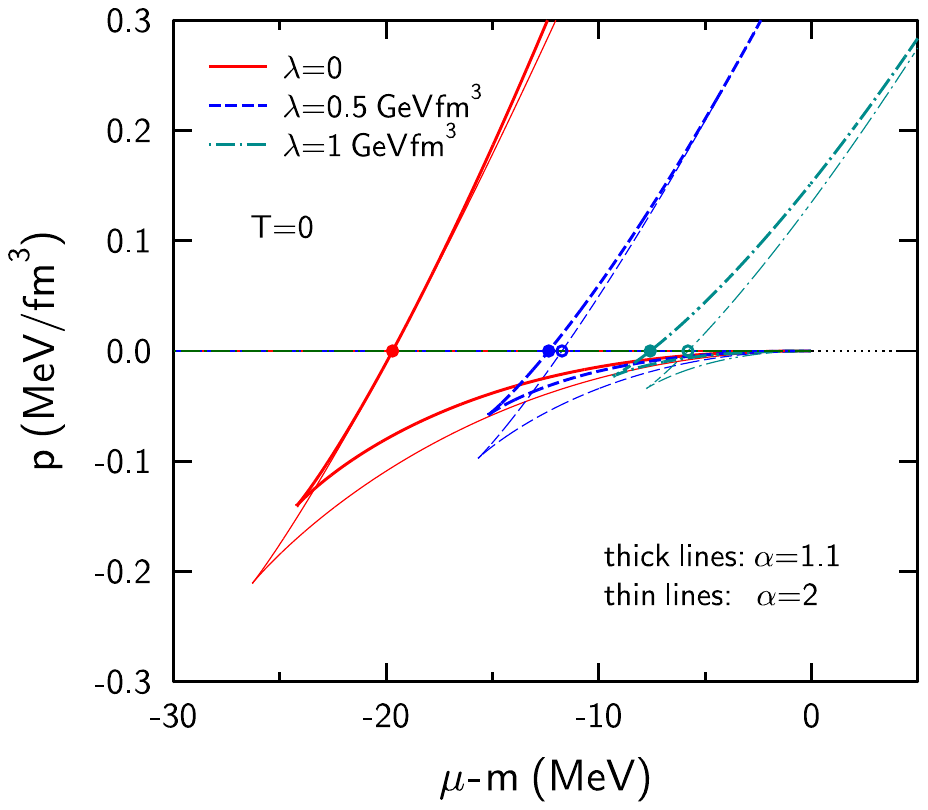}
\caption{
Pressure of cold BM as the function of chemical potential
at different values of parameter~$\lambda$\,.
Thick and thin lines are calculated with $\alpha=1.1$ and $\alpha=2$\hspm, respectively.
Full dots show positions of GS\,.
}\label{fig3}
\end{figure}
\begin{figure}[htb!]
\centering
\includegraphics[trim=2cm 7.5cm 3cm 8cm,width=0.6\textwidth]{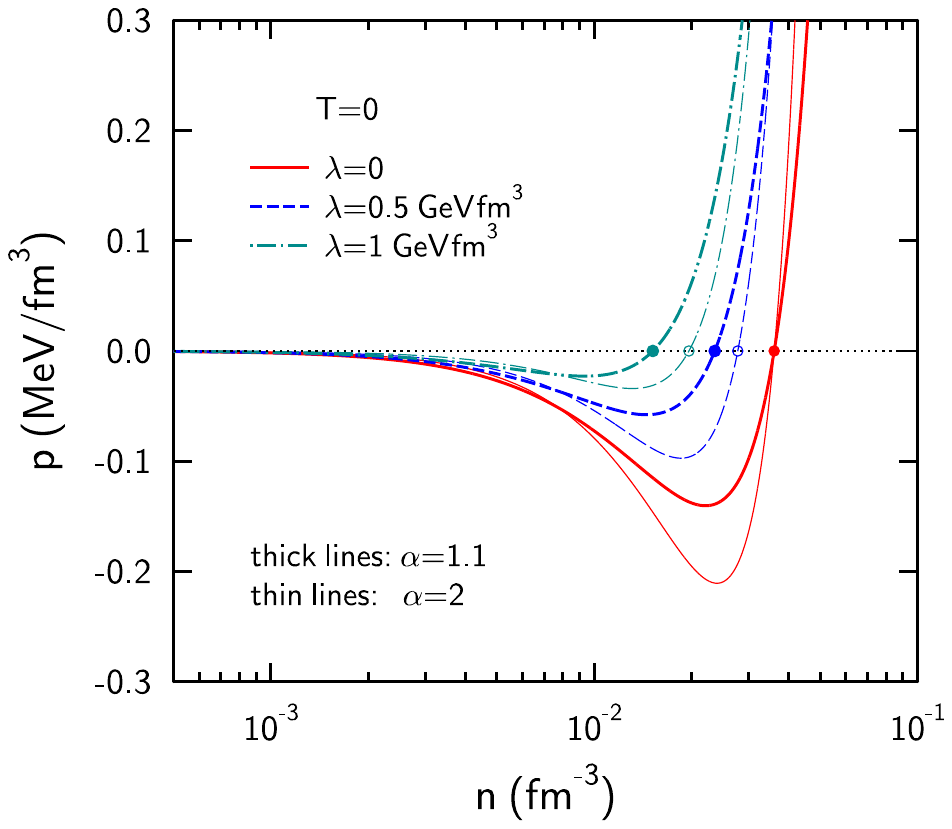}
\caption{
Same as Fig.~\ref{fig3}, but for pressure as the function of density.}
\label{fig4}
\end{figure}

A more detailed information is given in Figs.~\ref{fig1}--\ref{fig4}.
In particular, Fig.~\ref{fig1} show iso\-therms~\mbox{$T=0$} on the ($\mu,\sigma$) plane
for different parameters $\alpha$ and $\lambda$\,. At $\lambda=0$
the positions of~GS coincide for $\alpha=1.1$ and $\alpha=2$.
This follows from the procedure of choosing parameters of the Skyrme potential
used in present paper. One can see that each line
in Fig.~\ref{fig1} has upper and low parts. The low parts
with $\sigma^\prime\hspm (\mu)<0$ correspond to unstable (spinodal)
states with $c^2_s<0$. Within the present model, at $T=0$ the GS is
simultaneously the binodal point of the LGPT.
Note that pressure of BM changes sign at GS points.
The states with $\mu<\mu_0$ on the upper branch of $\sigma\hspm ({\mu})$
are in fact metastable\hsp\footnote
{
The metastable states with negative pressure have been observed
in liquid $^4$He~\cite{Pit00}.
}.

The same conclusions can be made from Fig.~~\ref{fig2}
which shows the sound velocity squared, calculated by
using \re{svel1}. The unstable spinodal states
correspond to points below the horizontal line $c_s^2=0$\,.
One can see that sound velocity vanishes at the boundary between metastable
and spinodal states. At this boundary the chemical potential
takes its minimal possible value.

The conclusion that GS is the binodal point of the LGPT can also be made
from Figs.~\ref{fig3}--\ref{fig4} which show the isoterms of pressure
on the $(\mu,p)$ and $(n,p)$ planes. The binodal states of LGPT originate
from intersecting pressure lines with small and large slopes in
Fig.~\ref{fig3}. This is a graphic representation of the Gibbs construction
rule for phase equilibrium between the liquid and gas phases (see Sec.~\ref{phd}).
At $T=0$ both phases have vanishing pressure. The spinodal 'states'
correspond to points between two kinks of pressure slopes in the $(\mu,p)$ plane.
For all considered values of $\lambda$ and $\alpha$ one gets the 'gas-like'
spinodal boundary at the same point with~$\mu=m$ and $p=0$\,.

\section{Thermodynamic functions at nonzero temperatures~\label{nzt}}

Below we modify the formalism, suggested in Ref.~\cite{Sat21} to describe
warm $\alpha$ matter without~DI. At temperatures $T\ne 0$,
in addition to spatially homogeneous BEC state $\phi_c$ with vanishing momentum, it is
necessary to take into account thermal fluctuations of scalar fields.
We use a plane-wave decomposition of $\phi$ in terms of creation ($a_{\bm{k}}^+$) and
annihilation~($a_{\bm{k}}$) operators~\cite{Lif80}
\bel{fdec}
\phi\hspm (x)=\phi_c\hsp (t)+\sum_{\bm{k}}\frac{1}{\sqrt{2 F E_k}}\left(a_{\bm{k}} e^{-ikx}+a_{\bm{k}}^+ e^{ikx}\right),
\ee
where \mbox{$\phi_c=\exp{(i\mu\hspm t)}\sqrt{\sigma_c}$} is proportional to the square root
of condensate scalar density $\sigma_c$\,, \mbox{$k^0=E_{\bm{k}}\equiv\sqrt{M^2+{\bm{k}}^2}$} ($M=M(\sigma)$ is the effective
mass of bosons defined in~\re{mum}). For brevity we introduce the notation
\bel{sumk}
\sum_{\bm{k}}=\frac{g}{(2\pi)^3}\int d^{\,3} k\,,
\ee
where $g$ is the spin-isospin statistical weight of a boson particle (\mbox{$g=1$} for alphas).
As~compared to Ref.~\cite{Sat21}, an additional factor $F$ is inserted into the denominator
of~\re{fdec}\hsp\footnote
{
This factor is introduced to preserve the standard equal-time commutation relation~\cite{Grei96}
$[\Pi\hspm (x),\phi(x')]=i\delta (\bm{r}-\bm{r^\prime})$, where \mbox{$\Pi\hspm (x)=\frac{\delta {\cal L}}{\delta {\dot\phi}^*(x)}=F\hspm \dot{\phi}/2$}
is the canonical momentum conjugated to $\phi$.
}.

In the following we denote by angular brackets the statistical averaging in the grand-canonical ensemble.
Within the mean-field approximation one gets the following expression for the particle momentum distribution
at given chemical potential~$\mu$ and temperature $T$:
\bel{pmd1}
n_{\bm{k}}=\langle a_{\bm{k}}^+a_{\bm{k}}\rangle=\left[\exp\left(\frac{E_{\bm{k}}-\mu}{T}\right)-1\right]^{-1}\,.
\ee
This distribution depends on the scalar density $\sigma=\langle\phi^2\rangle$ via the mass $M\hspm (\sigma)$ entering $E_{\bm{k}}$\,.
Note that allowed states of BM should satisfy the condition $\mu\leq M\hsp (\sigma)$ where the equal sign
corresponds to states with BEC. We shall see that this takes place at high enough $\mu$ or at temperatures
below the condensation temperature $T_{\rm BEC}(\mu)$\,.

Using Eqs.~(\ref{fdec}),~(\ref{pmd1}) one obtains the relation:
\bel{gape}
\sigma=\sigma_c+\sigma_{\rm th}\,,
\ee
where
\bel{sth}
\sigma_{\rm th}\hsp [T,\mu,M\hspm (\sigma)]=\sum_{\bm{k}}\frac{n_{\bm{k}}}{F E_{\bm{k}}}
=\frac{g}{2\hspm\pi^2F}\int\limits_M^\infty\frac{dE\sqrt{E^2-M^2}}{\exp{\left(\frac{\ds E-\mu}{\ds T}\right)-1}}\,.
\ee
The first condensate term in the right hand side of (\ref{gape})
vanishes for $T>T_{\rm BEC}$. In this case one obtains the self-consistent gap equation
$\sigma=\sigma_{\rm th}$ for $\sigma\hsp (T,\mu)$.
At $T<T_{\rm BEC}$ the scalar density is determined directly from the condition $M\hspm (\sigma)=\mu$.
Then \re{gape} may be regarded as the equation for determining $\sigma_c\hspm (T,\mu)$\,.

In the case $T\ll M$ one can use NRA by taking lower-order terms
in $T/M$ in calculating thermodynamic integrals. This leads to the approximate relation
\bel{snnr}
\sigma_{\rm th}\simeq\frac{g}{F M\lambda_T^3}\,g_{\,3/2}\hspm (z)\,,\\
\ee
where $z\leqslant 1$ is the nonrelativistic fugacity and $\lambda_T=\lambda_T\hsp (T,M)$ is the thermal wave length:
\bel{sntr1}
z=\exp\left(\frac{\mu-M}{T}\right),~~~~
\lambda_T=\sqrt{\frac{2\pi}{M\hsp T}}\,.
\ee
The dimensionless function (polylogarithm) $g_{\hspm\beta}\hspm (z)$ is defined as
\bel{bint}
g_{\hspm\beta}\hspm (z)\equiv\frac{1}{\Gamma(\beta)}\int\limits_0^\infty dx\hsp\frac{x^{\beta-1}}{z^{-1}e^{\ds x}-1}
=\sum_{k=1}^{\infty}z^k k^{-\beta}\,.
\ee
At low fugacities $g_{\hspm\beta}\hspm (z)\simeq z$ is small, in the degenerate limit $z\to 1$ one has $g_{\hspm\beta}(z)\simeq\xi (\beta)$ where $\xi (\beta)$
is the Riemann function.

By substituting (\ref{fdec}) into the first equality of (\ref{cen}) for $\nu=0$ and performing
the grand-canonical averaging one gets the relation for the particle number density:
\bel{dens2}
n=n_c+n_{\hsp\rm th}\,,
\ee
where $n_c=\mu F \sigma_c$ is  the Bose-condensed part of density and
\bel{dent}
n_{\hsp\rm th}\left(T,\mu,M\right)=\sum_{\bm{k}} n_{\bm{k}}=\frac{g}{2\hspm\pi^2}\int\limits_M^\infty\frac{dE\hspm E\sqrt{E^2-M^2}}{\exp{\left(\frac{\ds E-\mu}{\ds T}\right)}-1}\,.
\ee
Within the NRA one gets the approximate expression \mbox{$n_{\hspm\rm th}\simeq\frac{g}{\lambda_T^3}\,g_{\,3/2}\hspm (z)$}.
For states with \mbox{$n\lambda_T^3\gtrsim 1$}, the effects of quantum statistics are important.

One gets explicit
expression for the BEC boundary by substituting \mbox{$n_c=0, M=\mu$} in Eqs.~(\ref{dens2})--(\ref{dent}). One gets
the following formula for the threshold temperature of BEC in~NRA~\cite{Sat21}
\bel{tbec}
T_{\rm BEC}\approx\frac{2\pi}{\mu}\left(\frac{n}{g\hsp\xi\hspm (3/2)}\right)^{2/3}\,.
\ee

The pressure $p$ is determined by spatial components of the energy-momentum
tensor~$T_{\alpha\alpha}$ which in turn can be
calculated from the Lagrangian. One has in the mean--field approximation
\bel{pre1}
p=\frac{1}{3}\left<T_{xx}+T_{yy}+T_{zz}\right>=
p_{\,\rm ex}(\sigma)+p_{\hsp\rm th}\left(T,\mu,M\right)\,.
\ee
Here the first term is the so-called excess pressure
\bel{pre2}
p_{\,\rm ex}(\sigma)=\dfrac{M(\sigma)^2F\hsp\sigma}{2}-U(\sigma)\,.
\ee
It is equal to the pressure of cold homogeneous BM with
scalar density $\sigma$ (see~Eqs.~(\ref{den}) and~(\ref{pre0}))\hsp\footnote
{
This term vanishes for non-interacting BM with $a,b,\lambda\to 0$\,. Note that Bose-condensed
particles with zero momentum contribute the pressure only indirectly, via the
scalar density component $\sigma_c$\,.
}.
The second, thermal part takes the form
\bel{pre3}
p_{\,\rm th}\left(T,\mu,M\right)=\frac{F}{6}\left<{\bm
\nabla}\phi\hsp {\bm\nabla}\phi^*\hsp\right>=\sum_{\bm{k}}\frac{\bm k^2}{3E_{\bm{k}}}\hsp n_{\bm{k}}=
\frac{g}{6\pi^2}\int\limits_M^\infty\frac{dE\hspm \left(E^2-M^2\right)^{3/2}}{\exp{\left(\frac{\ds E-\mu}{\ds T}\right)}-1}\,.
\ee
In the second equality we applied the equation~(\ref{fdec}). The calculation shows that the excess pressure
dominates at small temperatures. One can see that expressions for~$n_{\,\rm th}$ and~$p_{\,\rm th}$ coincide
with the corresponding ideal gas expressions after the replacement \mbox{$m\to M$}  (note that factors~$F$ cancel out in Eqs. (\ref{dent}) and
(\ref{pre3})). Within the NRA one obtains the approximate
relation $p_{\,\rm th}\simeq\frac{g\hspm T}{\lambda_T^3}\,g_{\,5/2}\hspm (z)$\,.  In the Boltzmann
limit (at~$n\lambda_T^3\ll 1$) one has \mbox{$p_{\,\rm th}\simeq n_{\,\rm th}T$}.
One can check the validity of the relation $(\partial\hspm p_{\,\rm th}/\partial M)_{T,\hsp\mu}=-M\hspm F\hspm\sigma_{\rm th}$.
It~guarantees the thermodynamic consistency, $(\partial\hspm p/\partial \mu)_T=n$, of the present model~\cite{Sat21}.

\section{Phase diagram of bosonic matter\label{phd}}

\begin{figure}[htb!]
\centering
\includegraphics[trim=3cm 8cm 3cm 6.5cm,width=0.6\textwidth]{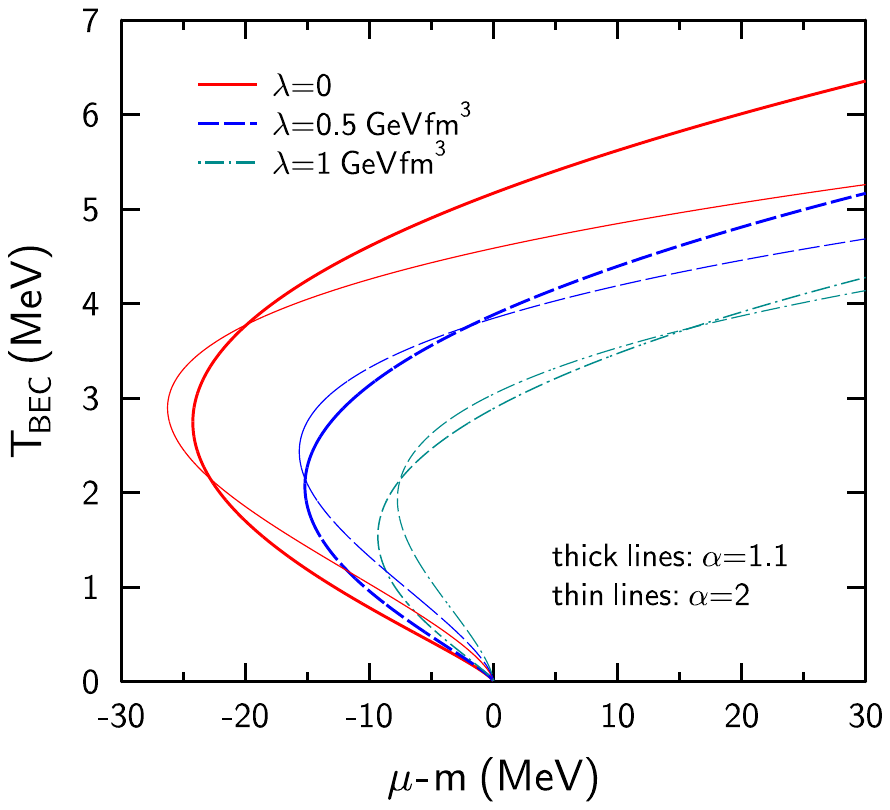}
\caption{
The BEC transition temperatures as functions of $\mu$. Thick and thin lines correspond
to $\alpha=1.1$ and $\alpha=2$ respectively.
}\label{fig5}
\end{figure}

Below we present the results of numerical calculations based on
the formulae of previous section. As in Sec.~\ref{numr}, we compare
the calculations for two choices of the Skyrme potential parameter
$\alpha=1.1$ and $\alpha=2$\,.
Figure~\ref{fig5} shows the BEC lines at various $\lambda$ and $\alpha$. Note that
lower (decreasing) parts of these lines are unstable.

\begin{figure}[htb!]
\centering
\includegraphics[trim=2cm 7.5cm 3cm 8cm,width=0.6\textwidth]{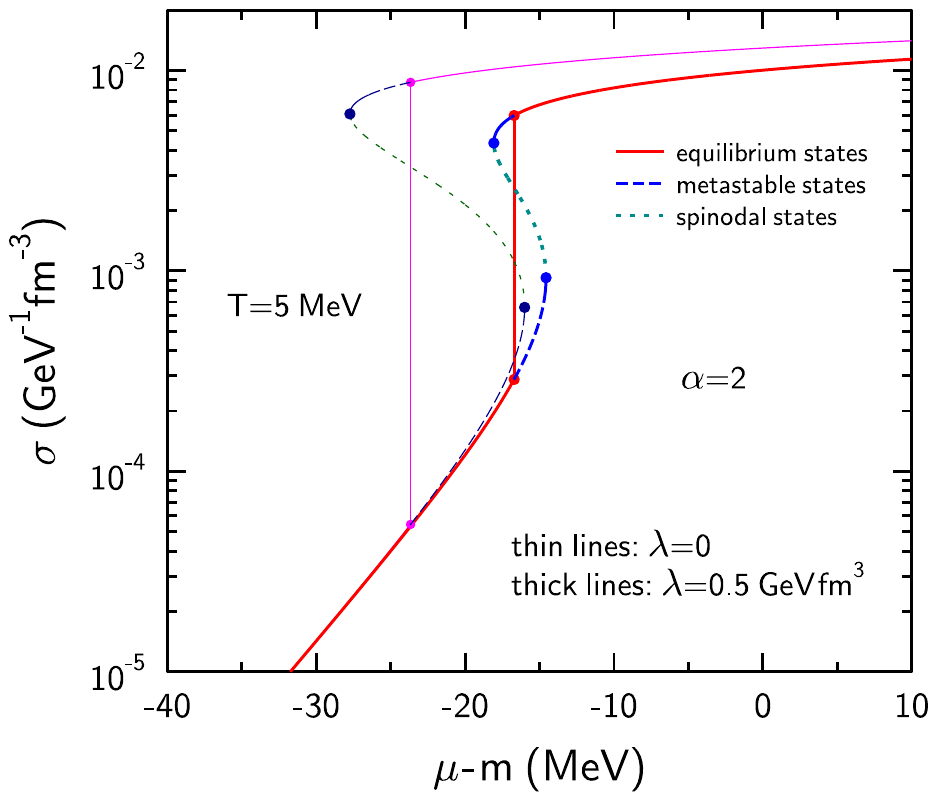}
\caption{The isotherms $T=5~\textrm{MeV}$ on the ($\mu,\sigma$) plane for $\alpha=2$\,. The vertical lines
correspond to equilibrium mixed-phase states of LGPT. The dashed and dotted curves represent
the metastable and spinodal (unstable) states, respectively.
}\label{fig6}
\end{figure}
To demonstrate the procedure of calculating phase diagrams we consider
in Fig.~\ref{fig6} the particular case of the isotherm $T=5~\textrm{MeV}$
for $\alpha=2$ and two values of $\lambda$\,. The vertical lines
in the $(\mu,\sigma)$ plane correspond to equilibrium mixed-phase states of the LGPT.
The upper and lower ends of these lines represent the liquid-like (L) and gas-like (G)
binodal states of~LGPT. The corresponding scalar densities $\sigma_i~(i=L,G)$ are found
from the Gibbs condition of the phase equilibrium
\begin{equation}
p\left[T,\mu,\sigma_L\hspm (T,\mu)\right]=p\left[T,\mu,\sigma_G\hspm (T,\mu)\right]~~~\label{gpt}\,.
\ee
Numerically we solve this equation by finding intersection of two branches
of isotherms with different slopes in the~($\mu,p$) plane. In this way
we determine the binodal line of LGPT as well as
the spinodal boundaries of LGPT. One can indeed see from Fig.~\ref{fig6} that the position
of the LGPT shifts to smaller $|\mu-m|$ with raising $\lambda$\,.

The critical point (CP) of the LGPT is the state with maximum temperature
where the solution of \re{gpt} still exists. The binodal scalar densities
coincide in this case:
\bel{cpc}
\sigma_L\hspm (T_c,\mu_c)=\sigma_G\hspm (T_c,\mu_c)\,.
\ee
The present model leads to characteristics of CP listed in
Tables \ref{tab3} and \ref{tab4}.
\begin{table}[ht!]
\caption
{\label{tab3} Characteristics of the critical points of LGPT
for $\alpha=1.1$\,.}
\vspace*{3mm}
\begin{tabular}{|c|c|c|c|c|}
\hline
~$\lambda\hsp\left(\textrm{GeV}\hspm\textrm{fm}^{3}\hsp\right)$~&~
$T_c\hsp\left(\textrm{MeV}\hsp\right)$~&
~$\mu_c-m\hsp\left(\textrm{MeV}\hsp\right)$~&~$n_c\times 10^2\hsp\left(\textrm{fm}^{-3}\right)$~&
~$p_c\times 10^2\hsp\left(\textrm{MeV}\hspm\textrm{fm}^{-3}\hsp\right)$~\\
\hline
~$0$~&~$9.90$~&~$-37.3$~&~$0.90$~&~$2.36$\\
\hline
~$0.5$~&~$6.30$~&~$-22.1$~&~$0.60$~&~$1.00$\\
\hline
~$1$~&~$3.95$~&~$-12.8$~&~$0.39$~&~$0.41$\\
\hline
\end{tabular}
\end{table}

\begin{table}[ht!]
\caption
{\label{tab4} Characteristics of the critical points of LGPT
for $\alpha=2$\,.}
\vspace*{3mm}
\begin{tabular}{|c|c|c|c|c|}
\hline
~$\lambda\hsp\left(\textrm{GeV}\hspm\textrm{fm}^{3}\hsp\right)$~&~
$T_c\hsp\left(\textrm{MeV}\hsp\right)$~&
~$\mu_c-m\hsp\left(\textrm{MeV}\hsp\right)$~&~$n_c\times 10^2\hsp\left(\textrm{fm}^{-3}\right)$~&
~$p_c\times 10^2\hsp\left(\textrm{MeV}\hspm\textrm{fm}^{-3}\hsp\right)$~\\
\hline
~$0$~&~$13.79$~&~$-49.0$~&~$1.24$~&~$5.73$\\
\hline
~$0.5$~&~$8.46$~&~$-25.9$~&~$0.97$~&~$2.75$\\
\hline
~$1$~&~$4.49$~&~$-10.7$~&~$0.71$~&~$1.06$\\
\hline
\end{tabular}
\end{table}

\begin{figure}[htb!]
\centering
\includegraphics[trim=2cm 9cm 3cm 7cm,width=0.6\textwidth]{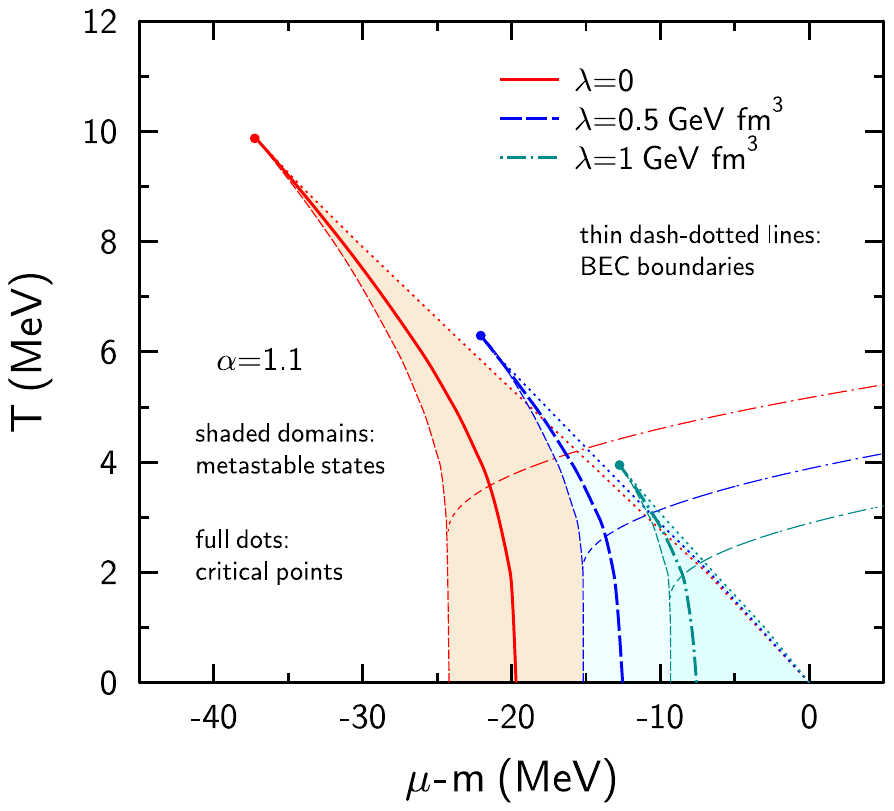}
\caption{Phase diagrams of bosonic matter on the ($\mu,T$) plane for $\alpha=1.1$ and different values
of the derivative strength $\lambda$. Thick lines correspond to mixed-phase states of the LGPT.
Shading show regions of metastable states. Dots represent positions of critical points.
The dash-dotted lines show boundaries of BEC states.
}\label{fig7}
\end{figure}
The phase diagrams of BM in the ($\mu,T$) plane predicted by our calculation
are shown in Figs.~\ref{fig7} and ~\ref{fig8} for $\alpha=1.1$ and $\alpha=2$,
respectively. In addition we show the BEC lines and boundaries of metastable regions.
The phase transition lines starts at zero-temperature GS points.
The gas-like spinodals for different $\lambda$ ends at the vacuum point $\mu=m, T=0$.
One can see that shifts of phase diagrams to smaller $T_c$ and $|\mu_c-m|$ occur with
increasing parameter~$\lambda$\,. Note that the BEC lines enter the
(meta)stable regions of LGPT at low enough $\mu$\,.
\begin{figure}[htb!]
\centering
\includegraphics[trim=2cm 9cm 3cm 7cm,width=0.6\textwidth]{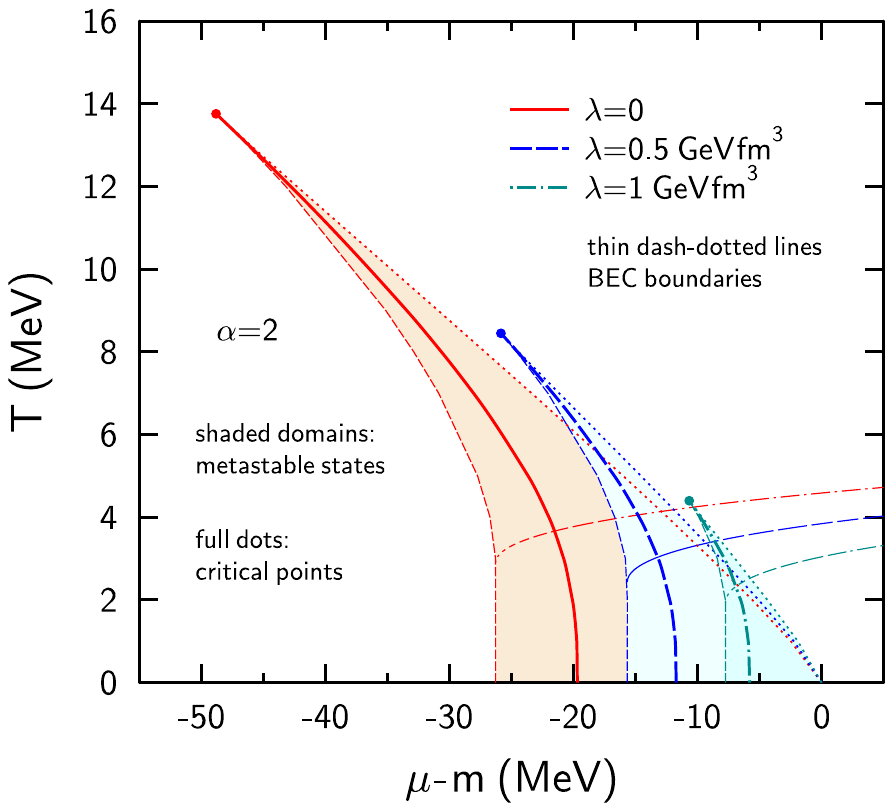}
\caption{
Same as Fig.~\ref{fig7}, but for $\alpha=2$\,.
}\label{fig8}
\end{figure}

The sensitivity of phase diagrams to the parameter $\alpha$ is demonstrated
in Fig.~\ref{fig9}.
\begin{figure}[htb!]
\centering
\includegraphics[trim=2cm 9cm 3cm 6.5cm,width=0.6\textwidth]{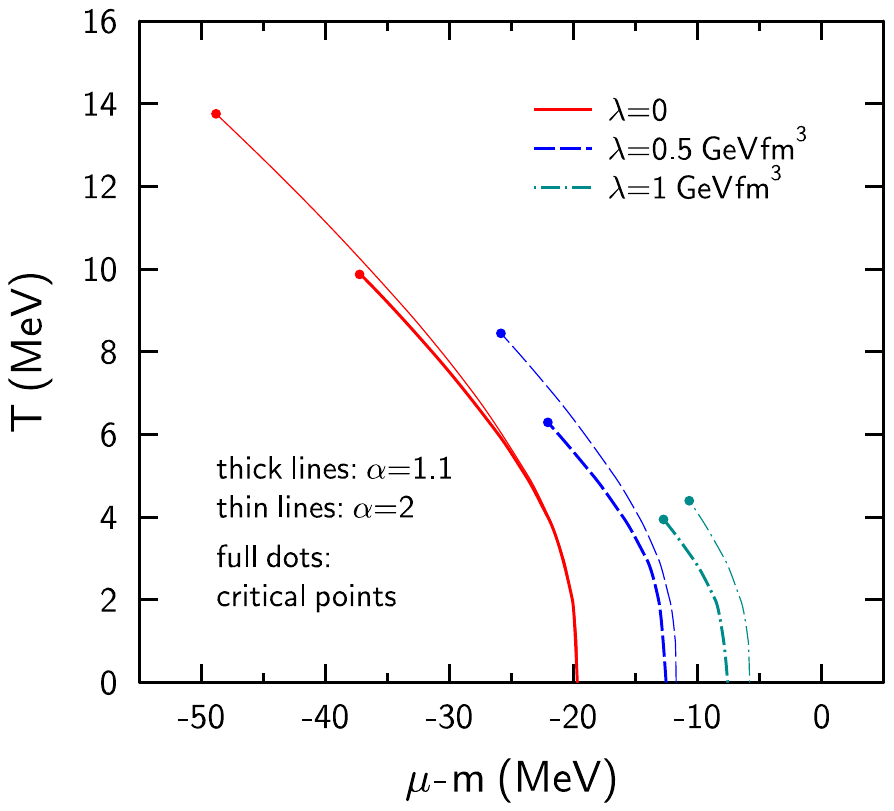}
\caption{Comparison of LGPT lines for $\alpha=1.1$ (thick) and $\alpha=2$ (thin)
on the ($\mu,T$) plane\,. Dots represent positions of critical points.
}\label{fig9}
\end{figure}

\section{Conclusions}

In the present paper we have investigated the sensitivity of the
phase diagram of bosonic matter to the derivative interaction. We assume that
the kinetic part of the Lagrangian contains the reduction factor $F$ diminishing
with the boson density. It is shown that this factor leads to an
additional repulsion which significantly modifies the equation of state
of bosonic matter as compared to the standard approach with $F=1$\,. We have
estimated  maximal strength of the derivative interaction when both the liquid-gas
phase transition and binding of bosonic matter disappear.

We think that the present approach will be useful for description
of bosonic systems in laboratory experiments, as well as for predicting
characteristics of boson stars and dark matter.

It would be interesting to extend current formalism by considering different parametrizations
of the derivative interaction. One can use the present model for studying properties of finite
bosonic systems (e.g. Q--balls). We believe that effects of the derivative interaction may be also
important for fermionic systems.

\begin{acknowledgments}
Authors appreciate the support from the Frankfurt
Institute for Advanced Studies.
\end{acknowledgments}


\end{document}